\documentclass{elsart}

\usepackage{graphics}
\usepackage{amssymb}
\begin{document}

\begin{frontmatter}

\title{Towards effective payoffs in the prisoner's dilemma game on scale-free networks}
\author[label1]{Attila Szolnoki \corauthref{cor1}},\ead{szolnoki@mfa.kfki.hu}
\author[label2]{Matja\v z Perc}, \ead{matjaz.perc@uni-mb.si}
\author[label3]{Zsuzsa Danku}
\address[label1]{Research Institute for Technical Physics and Materials Science, P.O. Box 49, H-1525 Budapest, Hungary}
\address[label2]{Department of Physics, Faculty of Natural Sciences and Mathematics, University of Maribor, Koro{\v s}ka cesta 160, SI-2000 Maribor, Slovenia}
\address[label3]{College of Ny{\' \i}regyh{\'a}za, H-4401, Hungary}
\corauth[cor1]{Corresponding author.}

\begin{abstract}
We study the transition towards effective payoffs in the prisoner's dilemma game on scale-free networks by introducing a normalization parameter guiding the system from accumulated payoffs to payoffs normalized with the connectivity of each agent. We show that during this transition the heterogeneity-based ability of scale-free networks to facilitate cooperative behavior deteriorates continuously, eventually collapsing with the results obtained on regular graphs. The strategy donations and adaptation probabilities of agents with different connectivities are studied. Results reveal that strategies generally spread from agents with larger towards agents with smaller degree. However, this strategy adoption flow reverses sharply in the fully normalized payoff limit. Surprisingly, cooperators occupy the hubs even if the averaged cooperation level due to partly normalized payoffs is moderate.
\end{abstract}

\end{frontmatter}

\section{Introduction}
\label{intro}

The welfare of a society often demands unselfish behavior of their members. Although working for the common good via cooperation seems a reasonable demand in return for a stable and flourishing society, it is often fundamentally at odds with the Darwinian principles of evolution suggesting that individuals should act exclusively so to enhance their own prosperity. While understanding the evolution of cooperation among selfish individuals is a puzzle faced by scientists across many different fields of research \cite{hammerstein_03}, the mathematical framework of choice for addressing the challenge is often the same, namely the evolutionary theory of games
\cite{weibull_95,hofbauer_98,gintis_00,nowak_06}. Particularly the prisoner's dilemma game, consisting of cooperators and defectors as the two possible strategies, is considered a paradigm for studying the emergence of cooperation between unrelated and selfish individuals \cite{axelrod_84}. The game promises a defecting individual the highest income if facing a cooperator. On the other hand, the cumulative income of two cooperators is higher than that of a cooperator-defector pair, and higher still than that of two defectors. According to the fundamental principles of Darwinian selection individuals should decide to defect, which ultimately results in social poverty. This unadorned scenario is indeed described by the classical well-mixed prisoner's dilemma game \cite{hofbauer_98}, where cooperators always die out. However, since cooperative behavior is an underpinning for a successful society, and is thus arguably present in everyday life 
\cite{milinski_n87,clutton_brock_s99}, the gap between the outlined theory and practice is obvious, indeed requiring refinement of the former. 

An important milestone in bridging the gap between the outcomes of the well-mixed prisoner's dilemma game and reality was the introduction of spatial structure and nearest neighbor interactions pioneered by Nowak and May \cite{nowak_n92}, which enabled the cooperators to form cluster on the spatial grid and so protect themselves against being exploited by defectors. The conditions of some biological experiments confirmed this conjecture \cite{kerr_n02}. However, there also exist compelling evidences that spatial structure may not necessarily favor cooperation \cite{hauert_n04}. Albeit a statement issued for the snowdrift game, in view of difficulties associated with the payoff rankings in experimental and field work \cite{milinski_prsb97,turner_n99}, it certainly carries an important message and dictates that the search for additional cooperation-facilitating mechanisms is necessary.

Indeed, research performed within recent years has made it clear that, besides spatial extensions and mechanisms such as kin selection, reciprocity, strategic complexity \cite{nowak_s06}, or asymmetry of learning and teaching activities \cite{szolnoki_epl07}, the specific topology of agents defining the interactions among them plays a key role in determining the outcome of the evolutionary process  \cite{abramson_pre01,santos_prl05,wu_pre05,santos_pnas06,perc_njp06,vukov_pre06,tang_epjb06,wu_pre06,chen_pa07,wu_pre07,oh_prl07}. Perhaps most prominently, scale-free (SF) networks \cite{barabasi_s99}, previously being identified as omnipresent in man made and natural systems \cite{albert_rmp02}, have been established as extremely potent promoters of cooperation in both the prisoner's dilemma as well as the snowdrift game \cite{santos_prl05,santos_pnas06}. In contrast, some works have shown that the ability of SF networks to promote cooperation disappears if we apply averaged instead of the more widely used accumulated payoffs
\cite{san_06,tomassini_p06,wu_pa07,masuda_prsb07}. Moreover, the interplay between the evolution of cooperation as well as that of the network defining the interactions among agents has also received considerable attention \cite{ebel_cm02,zimmermann_pre04,zimmermann_pre05,pacheco_prl06,sanplos_06,fu_p07}, showing specifically that the evolution of the interaction network might have a beneficial effect on cooperation, and finally, a widely applicable and simple rule for the evolution of cooperation on graphs and social networks has been proposed \cite{oh_na06}. For a comprehensive review of the field of research see \cite{szabo_pr07}.

In this paper, we wish to elaborate on the prominent role of scale-free networks and their ability to promote cooperation in the prisoner's dilemma game. More precisely, we study the transition towards effective payoffs in the prisoner's dilemma game on SF networks, whereby the latter is realized via the normalization of the payoffs of each agent with its connectivity. In particular, we introduce a normalization parameter that guides the system continuously from non-normalized, \textit{i.e.} absolute, to normalized, \textit{i.e.} effective, payoffs. A similar but technically different interpolation formula was recently proposed also by Tomassini \textit{et. al.} \cite{tomassini_p06}. However, our goal is not just to demonstrate deteriorated cooperator successfulness due to payoff normalization, but also to investigate the activity patterns of players that characterize the corresponding stationary states. For this purpose, we introduce strategy donation and adaptation probabilities of agents with different connectivities. These quantities enable us to rigorously identify the direction of the strategy flow. In other words, they serve to explore typical sources and targets characterizing the microscopic strategy adaptation process in the stationary state of the game. We find that cooperator densities of different classes of agents show a relevant dependence on their degree span, thus segregating the agents in terms of the underlying network topology. Moreover, we also study the global intensity of the adaptation process and reveal a resonant-type behavior depending on the payoff normalization parameter.

The paper is structured as follows. Section 2 is devoted to the description of the prisoner's dilemma game and the summarization of the main mechanism leading to an enhanced cooperation on SF networks. Also, it features the introduction of the normalization parameter that is being used to interpolate between the absolute and normalized payoffs. In Section 3 we present the results and discuss biological implications of our findings, whereas in the last Section we summarize our findings and conclude the paper.

\section{The Game}
\label{sec:game}

We consider an evolutionary two-strategy prisoner's dilemma game with agents located on vertices of a scale-free network generated via the celebrated mechanism of growth and preferential attachment \cite{barabasi_s99}, yielding a scale-free distribution of their connectivity $k$. Initially, each vertex $x$ is designated as a cooperator or defector with equal probability. Next, among all $k_x$ neighbors of a randomly chosen vertex $x$ one neighbor $y$ is also chosen at random, and both play one round of the prisoner's dilemma game with all their neighbors $k_x$ and $k_y$, respectively. Their accumulated payoffs resulting from $k_x$ and $k_y$ interactions are stored in $P_x$ and $P_y$. Thereby, the payoffs are determined in accordance with the standard prisoner's dilemma payoff matrix having temptation $b$, reward $1$, and both punishment as well as the suckers payoff $0$, where $1 < b < 2$ to ensure a proper payoff ranking. Crucially, before attempting strategy adoptions, the two accumulated payoffs are normalized according to
\begin{equation}
\overline{P_i} = \alpha P_i + (1-\alpha) { P_i \over k_i}
\label{eq:interpolate}
\end{equation}
where $i = x$ or $y$ and the newly introduced normalization parameter $\alpha$, occupying any value from the unit interval, determines exactly to what extent the payoffs are normalized with the number of interactions from which they were obtained. Evidently, $\alpha = 0$ dictates the use of fully normalized, so-called effective payoffs, whereas $\alpha = 1$ utilizes absolute, \textit{i.e.} accumulated, payoffs. Finally, in the spirit of the numerical implementation of the replicator dynamics, only if $\overline{P_y} > \overline{P_x}$ agent $x$ adopts the strategy of $y$ with probability $(\overline{P_y}-\overline{P_x})/[(1-\alpha)b + \alpha b k_m]$, where $k_m$ is the largest of $k_x$ and $k_y$. Since the strategy adoption is attempted only if $\overline{P_y} > \overline{P_x}$ this probability is always between $0$ and $1$, thus constituting the uncertainty related to each attempt of  the strategy adoption. An elementary part of the Monte Carlo (MC) algorithm, identical to the one often employed in population genetics, is thus given by first randomly choosing a pair of players and letting them both play the game with all their neighbors, and second, by attempting the strategy adoption only if $\overline{P_y} > \overline{P_x}$. In accordance with the random sequential update, each individual is selected once on average during a particular Monte Carlo step (MCS), which consists of repeating the elementary part $N$ times corresponding to all $x = 1, \dots, N$ vertices. Our simulations were performed on populations of $N = 5 \cdot 10^3$ to $5 \cdot 10^4$ agents, whereby the average connectivity of the scale-free network was always $z = 4$ as a result of the graph-generating algorithm \cite{barabasi_s99}. The application of the asynchronous MC update scheme diminishes the dependence of the stationary state on the initial distribution of the two strategies and the realization of the generated SF network \cite{tomassini_p06}. Characteristic quantities, such as the equilibrium frequencies of cooperators $\rho_C$ or the strategy adaptation probabilities defined below, were evaluated after sufficient transients $t_{tr} \approx 2 \cdot 10^4$ to $10^5$ MCS where the sampling period was two times longer than $t_{tr}$. All runs were repeated on $10$ to $30$ independently generated graphs. 

\begin{figure}
\resizebox{0.8\columnwidth}{!}{\includegraphics{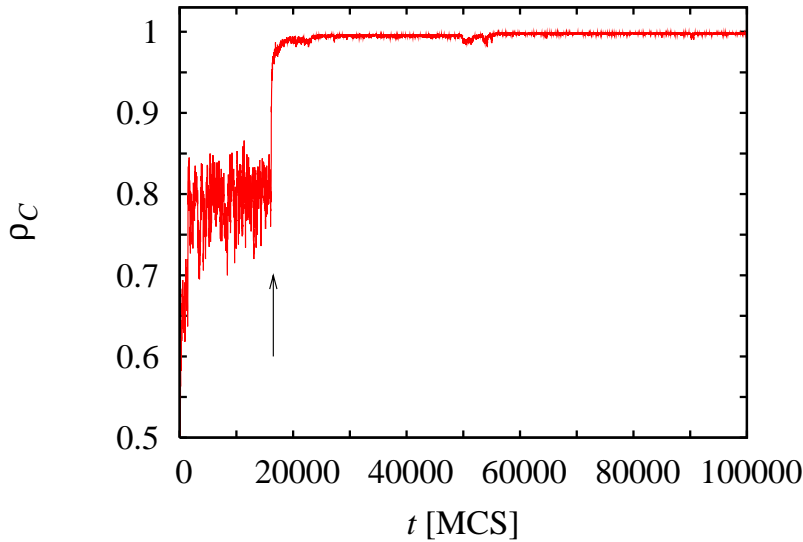}}
\caption{Temporal evolution of the cooperator density towards its stationary state for $b = 1.5$ and $\alpha = 1$. The arrow marks the time when the {\it last} defector hub was overtaken by a cooperator hub. After this event the whole defecting neighborhood of the defeated hub imitated the change, resulting in an enhanced level of cooperation.}
\label{fig:evol}
\end{figure}

We stress that our study focuses on the dynamical features of the game in its stationary state because the key mechanism leading to an enhanced cooperation during transient periods has already been uncovered in previous works \cite{szabo_pr07,santos_prsb06}. The most important topological feature of a SF network that promotes cooperation is the high connectedness of the hubs. Starting from a random distribution of strategies, all hubs can win over the neighbors because their hub-specific high connectivity results in large cumulative payoffs. Therefore the players that are linked to a hub will imitate its strategy, which eventually results in a cloud of homogeneous strategists around each hub \cite{gomez_gardenes_prl07}. In the long run, such an evolutionary process strengthens the cooperative and weakens the defector hubs. In particular, when two competing hubs play, which happens statistically later due to a small probability of such an event, the cooperating hub wins \cite{szabo_pr07}. Shortly afterwards, the whole cloud of defectors that was previously supported by the defeated defector hub adopts the cooperative strategy as well, which ultimately results in a substantial increase of cooperator density on the network. This mechanism is demonstrated in Fig.~\ref{fig:evol}, where the density of cooperators is plotted as a function of time at $b = 1.5$ warranting a high level of cooperation in the stationary state. At around $16000$ MCS (denoted by an arrow) the {\it last} defector hub changes its strategy, which results in a rapid increase of $\rho_c$ because shortly afterwards the whole cloud of defectors gives in to the cooperative strategy as well. The resulting state proves to be stable, bounding fluctuations of $\rho_c$ around a fixed steady-state value.

Before we turn to examining the results, some remarks regarding the motivation behind the introduction of the normalization parameter $\alpha$ are in order. In particular, while effective payoffs allow for a better mathematical study of the dynamics and system-size effects, as well as easier comparisons with analytical treatments both at the mean-field and pair-approximation level, on the other hand, absolute payoffs lead to a more exciting interplay of the two strategies as then cooperators can dominate not only by clustering (as on regular grids) but also via the benefits offered by the underlying SF interaction network. The dichotomy of results obtained via effective ($\alpha = 0$) and absolute ($\alpha = 1$) payoffs dictates the necessity for studying the behavior in-between the two extremes, and the explicit use of the interpolation formula given by Eq. (\ref{eq:interpolate}) enables an elegant accomplishment of this task. 

\section{Results}
\label{sec:3}

\begin{figure}
\resizebox{0.8\columnwidth}{!}{\includegraphics{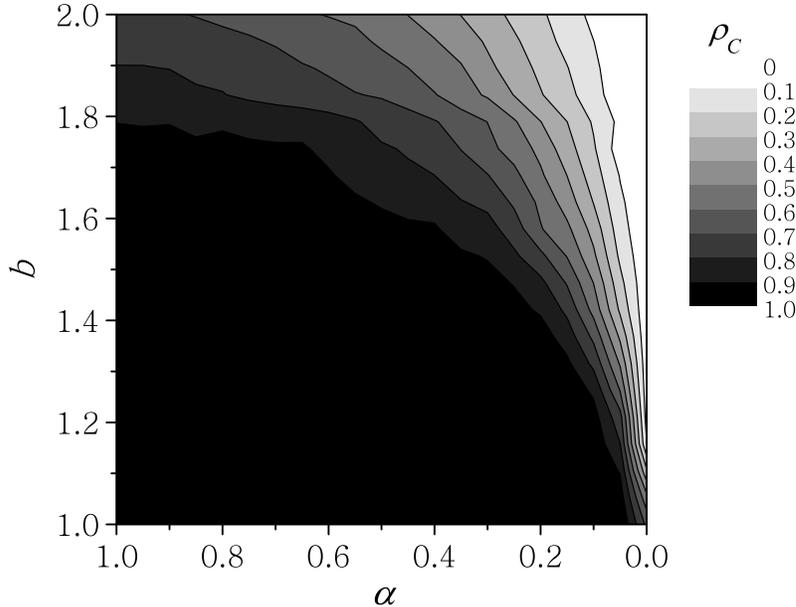}}
\caption{Fraction of cooperators $\rho_C$ in dependence on $b$ and the normalization parameter $\alpha$. Evidently, $\rho_C$ decreases continuously for all values of $b$ as $\alpha$ tends to $0$. Employed parameter values were: $N = 10^4$, $t_{tr} = 10^5$, and the results were averaged over $30$ different realizations of the underlying scale-free network.}
\label{fig:phd}
\end{figure}

We start by examining the impact of $\alpha$ on the equilibrium frequencies of cooperators and defectors. Figure~\ref{fig:phd} summarizes the results, from which it becomes instantly obvious that as $\alpha$ transits from $1$ towards $0$, thus gradually introducing the effective payoffs into the strategy adoption of the employed prisoner's dilemma game, the cooperative behavior suffers continuously. In particular, while the absolute payoffs ($\alpha = 1$) are able to sustain the domination of cooperation throughout the whole range of $b$, the effective payoffs ($\alpha = 0$) practically induce an extinction of the cooperative trait for all $b > 1.2$. In fact, results for $\alpha = 0$ are quite similar to the ones obtained for regular graphs with nearest neighbor interactions \cite{santos_prl05}, thus suggesting that the effective payoffs eliminate the advantage of cooperators given to them by the heterogeneity of the SF network \cite{wu_pa07}. We should mention that in \cite{tomassini_p06} a qualitatively identical result was reported, albeit by using a slightly different interpolation formula.

\begin{figure}
\resizebox{0.8\columnwidth}{!}{\includegraphics{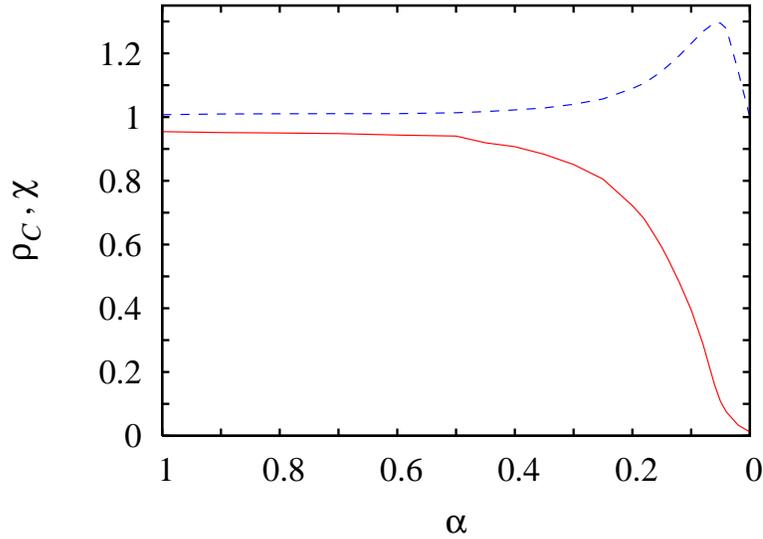}}
\caption{Ratio $\chi$ between the fractions of links that point to cooperators $\rho_{cl}$ and nodes occupied by cooperators $\rho_C$ in dependence on the normalization parameter $\alpha$ by $b = 1.5$ (dashed line). Solid line shows the decrease of the level of cooperation. These results suggest that hubs remain occupied by cooperators almost for all values of $\alpha$.}
\label{fig:cl}
\end{figure}

Although it seems very inviting to conclude that the normalization of payoffs via $\alpha$ simply eliminates the advantage of cooperator hubs given by their large number of links, arguably found crucial for the dissemination of cooperative behavior \cite{santos_prl05,szabo_pr07}, the dynamics behind results presented in Fig.~\ref{fig:phd} is in fact more subtle. To elaborate on this fact we first introduce the quantity $\chi = \rho_{cl} / \rho_C$, where $\rho_{cl}$ is the fraction of directed links ending at a cooperator. Although $\rho_{cl} \approx \rho_C$ by absorbing states, this is not necessarily true by mixed states. In particular, we argue that by mixed states $\chi$ provides complementary information about which strategy occupies the main hubs of the SF network. More precisely, $\rho_{cl} > \rho_C$ suggests that cooperators are residing on vertices with a larger number of links than defectors. Thus, values of $\chi > 1$ are a sufficient condition for the fact that hubs are predominantly occupied by the cooperative strategy. On the other hand, $\chi = 1$ indicates that the system is either in an absorbing state (which is trivially true), or more interestingly, that neither of the two strategies succeed in predominantly occupying the hubs of the network by mixed states. Results in Fig.~\ref{fig:cl} confirm this interpretation of $\chi$ since the value stays close to $1$ as long as the stationary state remains predominantly cooperative. Strikingly though, $\chi$ raises when $\rho_C$ decreases, signaling that the hubs remain predominantly occupied by cooperators despite the fact that values of $\alpha$ close to $0$ introduce effective payoffs that diminish the advantage of multiple connections. Naturally, when the majority of agents adopts the defecting strategy the hubs change their strategy as well, and thus $\chi$ again quickly converges to $1$. Nonetheless, results for values of $\alpha$ close to $0$ reveal that hubs are occupied by cooperators despite the apparent dismissal of their advantage that is implied by the normalization of the accumulated payoffs with the number of interaction from which they were obtained \cite{luthi_p07}.

\begin{figure}
\resizebox{0.8\columnwidth}{!}{\includegraphics{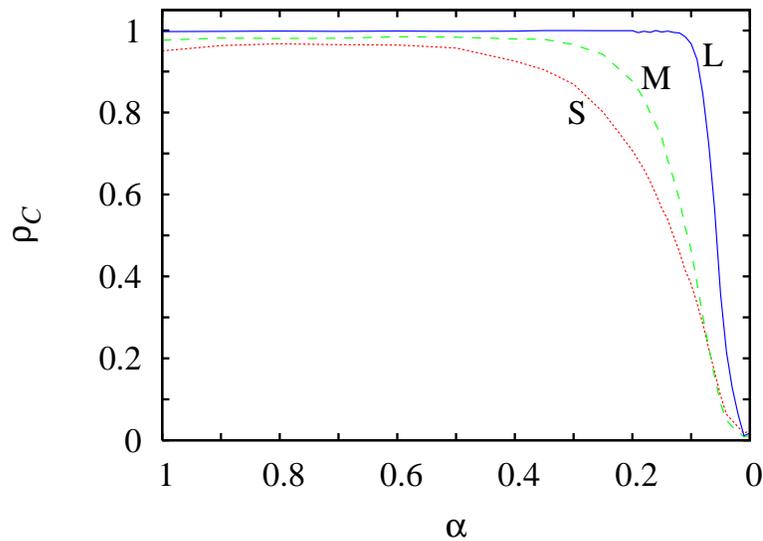}}
\caption{Fractions of nodes that are occupied by cooperators, depending on their connectivity $k$, as a function of $\alpha$ by $b = 1.5$. Results are presented separately for three connectivity classes, later being small (S - dotted line), medium (M - dashed line), and large (L - solid line). Cumulative values of $\rho_C$ displayed in Fig.~\ref{fig:cl} are close to the values depicted by the S curve.}
\label{fig:partc}
\end{figure}

Our conjecture can be additionally strengthened by studying $\rho_C$ for different classes of agents depending on their connectivity $k$. For this purpose, we divide the agents into three categories such that the connectivity span belonging to each group is equally large on the logarithmic scale. The fraction of cooperators in the group with the largest, medium, and the smallest connectivity is denoted by $\rho_C^L$, $\rho_C^M$, and $\rho_C^S$, respectively. Results presented in Fig.~\ref{fig:partc} clearly show that the main hubs are actually the ones that remain occupied by cooperators the longest as $\alpha$ decreases towards $0$, while in fact $\rho_C^M$, and even more so $\rho_C^S$, start to deteriorate substantially sooner. In sum, results presented in Figs.~\ref{fig:cl} and \ref{fig:partc} clearly demonstrate that, in contradiction with the intuitive reasoning, hubs remain occupied by cooperators until the actual $\alpha = 0$ limit is reached, despite the fact that even for values of $\alpha$ close to $0$ virtually effective payoffs are governing the strategy adoption process of the game. In other words, even a tiny benefit originating from the larger number of links suffices for cooperators to invade the hubs.

\begin{figure}
\resizebox{0.67\columnwidth}{!}{\includegraphics{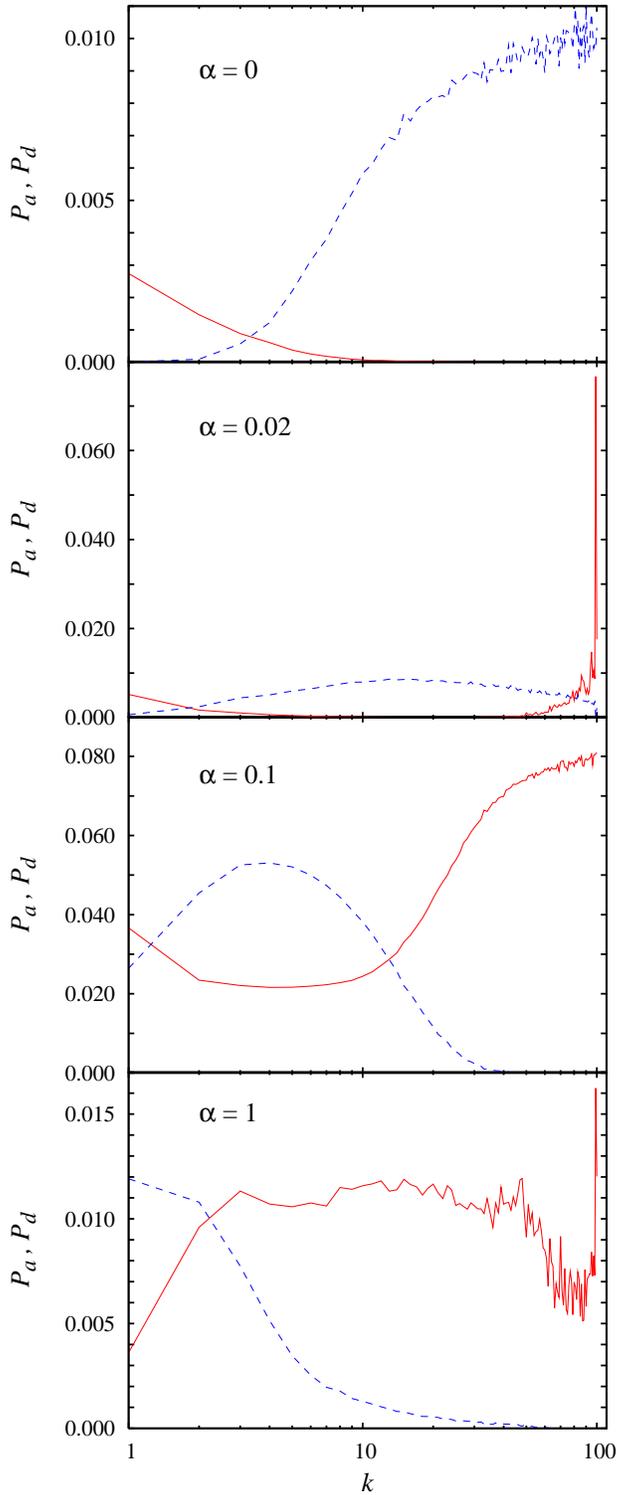}}
\caption{The probability of strategy donations $P_d$ (solid lines) and adaptations $P_a$ (dashed lines) in dependence on the connectivity of agents $k$. Values on the horizontal axis, running from $1$ to $100$, correspond to the minimal and maximal connectivity, respectively. The steady-state activity patterns are shown for four representative values of $\alpha$, all obtained by $b = 1.5$.}
\label{fig:multi}
\end{figure}

To get a deeper understanding of the presented results, we further investigate the microscopic mechanism behind the deterioration of the cooperation-facilitating effect of SF networks as $\alpha \rightarrow 0$. Therefore, we study the strategy adoption process of agents with different connectivity, thereby providing insights into the dynamics leading to the characteristic properties of stationary states. For this purpose we calculate the probability $P_a(k)$ that an agent with connectivity $k$ will adopt the strategy of its neighbor. This quantity characterizes the potential targets of the invasion process. To identify possible sources we also introduce $P_d(k)$, which is the probability that an agent with connectivity $k$ will pass (donate) the strategy to its neighbor during an elementary process. Figure~\ref{fig:multi} illustrates the introduced probability distributions for different values of $\alpha$. In case of accumulated payoffs ($\alpha = 1$) the strategies of agents with midddle and high connectivity are practically frozen. They cannot change but are only able to spread their strategies to agents that have fewer links. In the opposite limit ($\alpha = 0$), when effective payoffs apply, only agents with small connectivity can donate their strategy to other agents. The comparisons of $P_a$ and $P_d$ distributions reveal that the direction of strategy flow remains unidirectional for all $\alpha > 0$. The only exception thereby is the 
$\alpha = 0$ case. In sum, strategies generally spread from agents with larger to agents with smaller connectivity, and moreover, this strategy adoption flow reverses sharply only in the fully normalized payoff limit.

\begin{figure}
\resizebox{0.8\columnwidth}{!}{\includegraphics{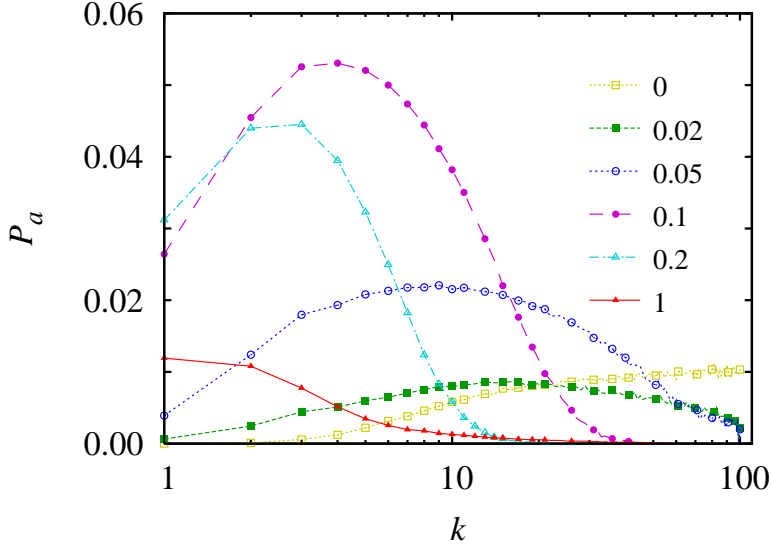}}
\caption{Comparative plots of strategy adaptation probabilities of agents with degree $k$ by different values of $\alpha$. Curves characterize possible targets of strategy adaptation.}
\label{fig:target}
\end{figure}

The comparison of "target distributions" obtained by different values of $\alpha$ presented in Fig.~\ref{fig:target} reveals that for $\alpha$ close to $0$ hubs change their strategy far more often than their counterparts by larger values of $\alpha$. As $\alpha$ increases towards 1, the bulk of adaptation activity shifts from agents with larger to agents with smaller connectivity via an intermediate non-monotonous phase. We argue that the mutable nature of cooperator hubs, observed for close to zero values of $\alpha$, is the main reason for their inability to spread cooperative behavior, despite the fact that, as suggested by results in Figs.~\ref{fig:cl} and \ref{fig:partc}, the majority of time defectors fail to conquer this prime spots of the network. Thus, the main mechanism behind the destructive impact of effective payoffs on the cooperative strategy is not the inability of cooperators to occupy the main hubs of the network, but rather their inability to permanently sustain that spots by comparatively (in comparison to results obtained for values of $\alpha$ close to $1$) frequently losing their battles to the defective intruders.

\begin{figure}
\resizebox{0.8\columnwidth}{!}{\includegraphics{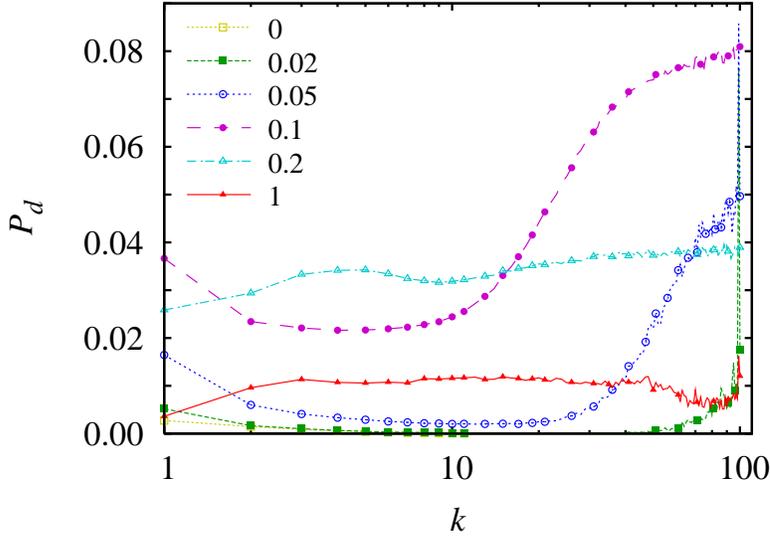}}
\caption{Comparative plots of strategy donation probabilities of agents with degree $k$ by different values of $\alpha$. Curves characterize possible sources of strategy adaptation.}
\label{fig:source}
\end{figure}

The target and source distributions presented in Figs.~\ref{fig:target} and \ref{fig:source} also reveal a resonant-type increase of the overall adaptation process as we change the value of $\alpha$ from $1$ to $0$. This non-monotonous curve, not shown here, can be well approximated in the mean-field level by the product of the two strategy densities $\rho_C \rho_D$. Consequently, in the coexistence phase the stationary state is not rigid but intensely fluctuates due to frequent strategy adaptations, especially when $\rho_C$ is far from the absorbing state.

\section{Summary}
\label{sec:x}

In sum, we elaborate on the prominent role of scale-free networks and their ability to promote cooperation in the prisoner's dilemma game. We study the transition towards effective payoffs that is driven via a single normalization parameter $\alpha$. We find that the main mechanism behind the destructive impact of effective payoffs on the cooperative strategy is not the inability of cooperators to occupy the main hubs of the network, but rather their inability to permanently sustain that spots by comparatively frequently losing their battles to the defective intruders. In case of normalized payoffs players having the smallest connectivity rule the evolution. The careful analysis of strategy donation and adaptation habits reveals that the flow of strategy transmissions reverses to the opposite direction instantly even if the payoffs governing the evolution shift only slightly from the effective towards the absolute case. We find also that the heterogeneous host network implies heterogeneity of the distribution of cooperators on nodes that have different connectivity, and moreover, that cooperators always prefer widely connected nodes (hubs) irrespectively of the level of payoff efficiency.

\begin{ack}
Discussions with Gy{\"o}rgy Szab{\'o} are gratefully acknowledged. This work was supported by the Hungarian National Research Fund (Grant No. T-47003) (A. S.) and the Slovenian Research Agency (Grant No. Z1-9629) (M. P.).
\end{ack}

\end{document}